\documentclass[12pt]{article}

\batchmode

\usepackage{amsfonts}

\newtheorem{theorem}{Theorem}

\newtheorem{definition}{Definition}

\def\={\mathop{\;=\;}}  

\newcounter{aid}

\begin{document}

\title{\bf A General Dependency Structure for Random Graphs and Its Effect on Monotone Properties}

\author{ \vspace*{1mm} Zohre Ranjbar-Mojaveri and Andr\'as  Farag\'o \\
Department of Computer  Science      \\
The  University   of  Texas   at  Dallas\\
   Richardson,  Texas \\
 }

\date{}
\maketitle

\begin{abstract}
We consider random graphs in which the edges are allowed to be {\em dependent}. 
In our model the edge dependence is quite general, we call it $p$-robust random graph. It means that every edge is present with probability at least $p$,
regardless of the presence/absence of other edges. This is more general than independent edges with probability $p$, as we illustrate with examples.
Our main result is that for {\em any monotone graph property,} the $p$-robust random graph has at least as high probability to have the property as an 
Erd\H os-R\'enyi random graph with edge probability $p$. This is very useful, as it allows the adaptation of many results from classical Erd\H os-R\'enyi random graphs to a non-independent setting, as lower bounds.
\end{abstract}
  
\section{Introduction}

Random networks occur in many practical scenarios. Some examples are wireless ad-hoc networks, various social networks, the web graph describing the 
World Wide Web, and a multitude of others.  Random graph models are often used to describe and analyze such networks.

The oldest and most researched random graph model is the Erd\H os-R\'enyi random graph $G_{n,p}$. This denotes a random graph on $n$ nodes, 
such that each edge is added with probability $p$, and it is done {\em independently} for each edge. 
A large number of  deep results are available about such random graphs, see expositions in the books \cite{bollobas,frieze,janson}. Below we list some examples.
They are asymptotic results, and for simplicity/clarity we omit potential restrictions for the range of $p$, as well as ignore rounding issues (i.e., an asymptotic  formula may provide a non-integer value for a parameter which is defined as integer for finite graphs).
\begin{itemize}

\item The size of a maximum clique in $G_{n,p}$ is asymptotically $2\log_{1/p} n$.

\item If $G_{n,p}$ has average degree $d$, then its maximum independent set has asymptotic size $\frac{2n\ln d}{d}$.

\item The chromatic number of $G_{n,p}$ is asymptotically $\frac{n}{\log_b n}$, where $b=\frac{1}{1-p}$.

\item The size of a minimum dominating set in $G_{n,p}$ is asymptotically $\log_b n$, where $b=\frac{1}{1-p}$.

\item The length of the longest cycle in $G_{n,p}$, when the graph has a constant average degree $d$, is asymptotically $n(1-d{\rm e}^{-d})$.

\item The diameter of $G_{n,p}$ is asymptotically $\frac{\log n}{\log (np)}$, when $np\rightarrow\infty$. (If the graph is not connected, then the 
diameter is defined as the largest diameter of its connected components.)

\item If $G_{n,p}$ has average degree $d$, then the number of nodes of degree $k$ in the graph is asymptotically $\frac{d^k {\rm e}^{-d}}{k!}n$.

\end{itemize}

These results (and many others) make it possible that for random graphs that one can find good and directly computable estimates of graph parameters that are hard to compute for deterministic graphs. Moreover, the parameters often show very strong concentration. For example, as listed above, the 
chromatic number of $G_{n,p}$ is asymptotically $\frac{n}{\log_b n}$, where $b=\frac{1}{1-p}$. However, we can say more: the chromatic number of a 
random graph is so strongly concentrated that with probability approaching one, as $n\rightarrow\infty$, it falls on one of two consecutive integers (see Alon and 
Krivelevich \cite{alon}).

\section{Random Graphs With Dependent Edges}

The requirement that the edges are {\em independent} is often a severe restriction in modeling real-life networks.
Therefore, numerous attempts have been made to develop models with various dependencies among the edges, see, e.g.,  a survey in \cite{farago}.
Here we consider a general form of edge dependency. We call a random graph with this type of dependency a {\em $p$-robust random graph.}

\begin{definition}{\bf ($p$-robust random graph)}
A random graph on $n$ vertices is called {\em $p$-robust}, if every edge is present with probability at least $p$, regardless of the 
status (present or not) of other edges. Such a random graph is denoted by $\widetilde G_{n,p}$.
\end{definition}

Note that $p$-robustness does not imply independence. It allows that the existence probability of an edge may depend on other edges, 
possibly in a complicated way, 
it only requires that the probability never drops below $p$. 
Let us show some examples of $p$-robust random graphs.
\begin{description}

\item[Example 1] First note that the classical Erd\H os-R\'enyi random graph $G_{n,p}$ is a special case of $\widetilde G_{n,p}$, since our model also allows adding all edges independently with probability $p$.

\item[Example 2]
However, we can also allow (possibly messy) dependencies. 
For example, let $P(e)$ denote the probability that a given edge $e$ is present in the graph, and let us condition on $k$, the number 
of {\em other} edges in the whole graph. Let $P_e(k)$ denote the probability that there are $k$ edges in the graph, other than $e$. 
For any fixed $k$, set $P(e|k)= 1-(k+1)/n^2$; let this be the probability that edge $e$ exists, given 
that there are $k$ other edges in the graph. Using that the total number of edges cannot be more than $n(n-1)/2$, we have that $k\leq n(n-1)/2-1$ always holds. Therefore, 
$P(e|k)\geq 1-\frac{n(n-1)}{2n^2} = 1-\frac{n-1}{2n}\geq \frac{1}{2}$, for any  $k$, implying 
$P(e)=\sum_{k=0}^{n(n-1)/2-1} P(e|k)P_e(k)\geq 1/2$. Thus, with $p=1/2$, this random graph is $p$-robust. At the same time, the edges are not independent, since the probability that $e$ is present depends on how many other edges are present. 

\item[Example 3]  For a given edge $e$, let $r(e)$ denote the number of edges that are adjacent with $e$ (not including $e$ itself). If $e$ does not exist, then let $r(e)=0$.
Let the conditional probability that
edge $e$ exists, given that it has $k$ adjacent edges, be $P(e|r(e)=k)=\frac{1}{2}-\frac{1}{k+5}$. Note that the possible range of $k$  
is $0\leq k\leq 2(n-2)$. Then we have
$P(e|r(e)=k)\geq \frac{1}{2}-\frac{1}{5}=\frac{3}{10}$. This implies 
$P(e)=\sum_{k=0}^{2(n-2)} P(e|r(e)=k)P(r(e)=k)\geq \frac{3}{10} \sum_{k=0}^{2(n-2)} P(r(e)=k)=\frac{3}{10}$.
Thus, with $p=\frac{3}{10}$, this random graph is $p$-robust. At the same time, the edges are not independent, since the probability that $e$ is present is influenced by the number of adjacent edges.

\item[Example 4] Consider the model described above in 3, but with the additional condition that each potential edge $e$ has at least 3 adjacent edges,
whether or not $e$ is in the graph. What can we say about this conditional random graph? The same derivation as in 3, but with $k\geq 3$,
gives us that the new random graph will remain $p$-robust, but now  with $p=\frac{3}{8}$.

\end{description}

If we have a random graph like the ones in Examples 2,3,4 above (and many possible others with dependent edges), then how can we estimate some parameter of the random graph, like the size of the maximum clique? We show that at least for so called monotone properties we can use the existing results about  
Erd\H os-R\'enyi random graphs as lower bounds.

Let $Q$ be a set of graphs. We use it to represent a graph property: a graph $G$ has property $Q$ if and only if $G\in Q$. Therefore, we
identify the property with $Q$. We are going to consider {\em monotone graph properties,} as defined below.

\begin{definition}
{\bf (Monotone graph property)}
A graph property $Q$ is called {\em monotone,} if it is closed with respect to adding new edges. That is, if $G\in Q$ and $G\subseteq G'$,
then $G'\in Q$. 
\end{definition}

Note that many important graph properties are monotone. 
Examples: having a clique of size at least $k$, having a Hamiltonian circuit, having $k$ disjoint spanning trees, having chromatic number at least $k$, 
having diameter at most $k$, having a dominating set of size at most $k$, having a matching of size at least $k$, and numerous others.
In fact, almost all interesting graph properties have a monotone version.
Our result is that for any monotone graph property, and for any $n,p$, it always holds that $\widetilde G_{n,p}$ is more likely to 
have the property than $G_{n,p}$ (or at least as likely). This is very useful, as it allows the application of the rich treasury of results
on Erd\H os-R\'enyi random graphs to the non-independent setting, as lower bounds on the probability of having a monotone property.
Below we state and prove our general result.

\begin{theorem}\label{p-robgr}
Let $Q$ be any monotone graph property. Then the following holds:
$$\Pr(G_{n,p}\in Q)\;\leq \; \Pr(\widetilde G_{n,p}\in Q).$$
\end{theorem}

\noindent 
{\bf Proof.} 
We are going to generate  $\widetilde G_{n,p}$ as the union of two random graphs, $G_{n,p}$ and $G_2$, both on the same vertex set $V$. 
$G_{n,p}$ is the usual Erd\H os-R\'enyi random graph, $G_2$ will be defined later.
The union $G_{n,p}\cup G_2$ is meant with the understanding that if the same edge occurs  in both graphs, then we merge 
them into a single edge. We plan to chose the edge probabilities in $G_2$,
such that $G_{n,p}\cup G_2\sim \widetilde G_{n,p}$, where the ``$\sim$" relation between random graphs means that they have the same 
distribution, i.e., they are statistically indistinguishable.
If this can be accomplished, then the claim will directly follow, 
since then a random graph distributed as $\widetilde G_{n,p}$ can be obtained by adding edges to $G_{n,p}$, which cannot destroy a monotone property,
once $G_{n,p}$ has it.  This will imply the claim.

We introduce some notations. Let $e_1,\ldots,e_m$ denote the (potential) edges.
For every $i$, let $h_i$ be the indicator of the event that the edge $e_i$ is included in  $\widetilde G_{n,p}$. 
Further, let us use the abbreviation $h_i^m=(h_i,\ldots,h_m)$.  
For any $a=(a_1,\ldots,a_m)\in \{0,1\}^m$, the event 
$\{h_1^m=a\}$ means that $\widetilde G_{n,p}$ takes a realization 
in which edge $e_i$ is included if and only $a_i=1$. Similarly, $\{h_i^m=a_i^m\}$ means $\{h_i=a_i,\ldots,h_m=a_m\}$.
We also use the abbreviation $a_i^m=(a_i,\ldots,a_m)$.  
Now let us generate the random graphs $G_{n,p}$ and $G_2$, as follows. 
\begin{description}

\item[\hspace*{5mm} \rm Step 1.] 
Let $i=m$.

\item[\hspace*{5mm} \rm Step 2.] 
If $i=m$, then let $q_m=\Pr(h_m=1)$.
If $i<m$, then set $q_i=\Pr(h_i=1\;|\; h_{i+1}^m=a_{i+1}^m)$, where  
$a_{i+1}^m$ indicates the already generated edges of $G_{n,p}\cup G_2$.

\item[\hspace*{5mm} \rm Step 3.]
Compute  
\begin{equation}\label{p'_i}
p'_i=\frac{p(1-q_i)}{1-p}.
\end{equation}

\item[\hspace*{5mm} \rm Step 4.] 
Put $e_i$ into $G_{n,p}$ with probability $p$, and put $e_i$ into $G_2$ with probability $q_i-p'_i$.

\item[\hspace*{5mm} \rm Step 5.]
If $i>1$, then decrease $i$ by one, and go to Step 2; else {\sc halt.}

\end{description}
First note that the value $q_i-p'_i$ in Step 4 can indeed be used as a probability. Clearly, $q_i-p'_i\leq 1$ holds, 
as   $q_i$ is a probability and $p'_i\geq 0$. To show $q_i-p'_i\geq 0$, observe that 
$ 
p'_i =\frac{p(1-q_i)}{1-p} \leq q_i,
$
since the inequality can be rearranged into $p(1-q_i)\leq q_i(1-p)$, which simplifies to $p\leq q_i$. The latter is indeed true, due to 
$q_i=\Pr(h_i=1\;|\; h_{i+1}^m=a_{i+1}^m)\geq p$,
which follows from the $p$-robust property.

Next we show that the algorithm generates the random graphs $G_{n,p}$ and $G_2$ in a way  that they satisfy  $G_{n,p}\cup G_2\sim \widetilde G_{n,p}$. We prove it by induction, starting from $i=m$ and progressing downward to $i=1$. For any $i$, let $G^i_{n,p}$, $G^i_2$ 
denote the already generated parts of $G_{n,p}, G_2$, respectively, after  executing Step 4 \, $m-i+1$ times, so they can only contain edges
with index $\geq i$. Further, let $\widetilde G^i_{n,p}$ be the subgraph
of $\widetilde G_{n,p}$ in which we only keep the edges with index $\geq i$, that is, $\widetilde G^i_{n,p}=\widetilde G_{n,p}-\{e_{i-1},\ldots,e_1\}$.
The inductive proof will show that 
$G^i_{n,p}\cup G^i_2\sim \widetilde G^i_{n,p}$ holds for every $i$.
At the end of the induction, having reached $i=1$, we are going to get  $G^1_{n,p} \cup G^1_2 \sim \widetilde G^1_{n,p}$,
which is the same as $G_{n,p} \cup G_2 \sim \widetilde G_{n,p}$.

Let us consider first the base case  $i=m$. Then we have $\Pr(e_m\in G_{n,p})= \Pr(e_m\in G^m_{n,p})=p$ by Step 4. 
Then in Step 4, edge $e_m$ is put into $G_2$ with probability
$q_m-p'_m$, yielding $\Pr(e_m\in G^m_2)=q_m-p'_m$.
Now observe that the formula (\ref{p'_i}) is chosen such that $p'_i$ is precisely the solution of the equation
\begin{equation}\label{p'eq}
p+ q_i-p'_i -(q_i-p'_i)p = q_i
\end{equation}
for $p'_i$. 
For $i=m$ the   equation becomes
\begin{equation}\label{p'eqm}
p+ q_m-p'_m  -(q_m-p'_m)p = q_m,
\end{equation}
and $p'_m=\frac{p(1-q_m)}{1-p}$ is the solution of this equation. Since by Step 4 we have $\Pr(e_m\in G^m_{n,p})=p$ and $\Pr(e_m\in G^m_2)=q_m-p'_m$, therefore,
we get that the left-hand side of (\ref{p'eqm}) is precisely the probability of the event $\{e_m\in G^m_{n,p}\cup G^m_2\}$.
By (\ref{p'eqm}), this probability is equal to $q_m$, which is set to $q_m=\Pr(h_m=1)=\Pr(e_m\in \widetilde G^m_{n,p})$ in Step 2.
This means that $G^m_{n,p}\cup G^m_2\sim \widetilde G^m_{n,p}$, as desired.

For the induction step, assume that the claim is true for $i+1$, i.e., $G^{i+1}_{n,p}\cup G^{i+1}_2\sim \widetilde G^{i+1}_{n,p}$ holds.
In Step 4, edge $e_i$ is added to $G^{i+1}_{n,p}$ with probability $p$. It is also added to $G^{i+1}_2$ with probability 
$q_i-p'_i$.
 Therefore, just like in the base case, we get that 
$p+q_i-p'_i -(q_i-p'_i)p=\Pr(e_i\in G^i_{n,p}\cup G^i_2).$
We already know that $p'_i$ satisfies the equation (\ref{p'eq}), so $e_i$ is added to $\widetilde G^{i+1}_{n,p}$ with probability 
$q_i=\Pr(h_i=1\;|\; h_{i+1}^m=a_{i+1}^m)$, given the already generated part,
represented by $a_{i+1}^m$. By the inductive assumption, $h_{i+1}^m$ is distributed as $\widetilde G^{i+1}_{n,p}$, 
which is the truncated version of $\widetilde G_{n,p}$, keeping only the $\geq i+1$ indexed edges. 
Hence, for  $h_{i+1}^m$, we can write by the chain rule of conditional probabilities:
$$ \Pr(h_{i+1}^m=a_{i+1}^m)=
\Pr(h_m=a_m)\prod_{j=i+1}^{m-1} \Pr(h_j=a_j\;|\; h_{j+1}^m=a_{j+1}^m).
$$
After processing $e_i$ (i.e., adding it with probability $q_i$), we get 
\begin{eqnarray}\nonumber
\Pr(h_i^m=a_i^m) &=& \Pr(h_i=a_i\;|\; h_{i+1}^m=a_{i+1}^m) \Pr(h_{i+1}^m=a_{i+1}^m) \\ \nonumber
&=& \Pr(h_i=a_i\;|\; h_{i+1}^m=a_{i+1}^m) 
\underbrace{\Pr(h_m=a_m)\prod_{j=i+1}^{m-1} \Pr(h_j=a_j\;|\; h_{j+1}^m=a_{j+1}^m)}_{\Pr(h_{i+1}^m=a_{i+1}^m)}\\ \nonumber
&=& \Pr(h_m=a_m)\prod_{j=i}^{m-1} \Pr(h_j=a_j\;|\; h_{j+1}^m=a_{j+1}^m), 
\end{eqnarray}
which, by the chain rule, is indeed the distribution of $\widetilde G^i_{n,p}$, completing the induction.

Thus, at the end, a realization $a=a_1^m\in \{0,1\}^m$ of $\widetilde G_{n,p}$ is generated with probability 
$$
\Pr(h_1^m=a) = \Pr(h_m=a_m)\prod_{j=1}^{m-1} \Pr(h_j=a_j\;|\; h_{j+1}^m=a_{j+1}^m),
$$
indeed creating $\widetilde G_{n,p}$ with its correct probability. Therefore,  we get 
$G_{n,p} \cup G_2 \sim \widetilde G_{n,p}$, so $\widetilde G_{n,p}$ arises by adding 
edges to $G_{n,p}$, which cannot destroy a monotone property. This implies the statement of the Theorem, completing the proof.

\hspace*{10mm} \hfill $\spadesuit$

\section{An Example}

As a sample application of the result, consider the random graph described in Example 3. For handy access, let us repeat the example here:
\begin{quote}
{\bf Example 3.}
For a given edge $e$, let $r(e)$ denote the number of edges that are adjacent with $e$ (not including $e$ itself). If $e$ does not exist, then let $r(e)=0$.
Let the conditional probability that
edge $e$ exists, given that it has $k$ adjacent edges, be $$P(e|r(e)=k)=\frac{1}{2}-\frac{1}{k+5}.$$ Note that the possible range of $k$  
is $0\leq k\leq 2(n-2)$. Then we have
$$P(e|r(e)=k)\geq \frac{1}{2}-\frac{1}{5}=\frac{3}{10}.$$ This implies 
$$P(e)=\sum_{k=0}^{2(n-2)} P(e|r(e)=k)P(r(e)=k)\;\geq\; \frac{3}{10} \sum_{k=0}^{2(n-2)} P(r(e)=k)=\frac{3}{10},$$
Thus, with $p=\frac{3}{10}$, this random graph is $p$-robust. At the same time, the edges are not independent, since the probability that $e$ is present is influenced by the number of adjacent edges.
\end{quote}

Now we can claim that this random graph asymptotically has a maximum clique of size 
 at least $2\log_{10/3}n$, chromatic number at least 
 $\frac{n}{\log_b n}$ 
 with 
 $b=\frac{1}{1-p}=\frac{10}{7}$, 
 minimum dominating set at most $\log_b n$, 
also with $b=10/7$, and diameter at most $\frac{\log n}{\log (3n/10)}$ .
All these follow from the known results about the Erd\H os-R\'enyi random graph $G_{n,p}$ (listed in the Introduction), complemented with our Theorem 1.
Note that such bounds would be very hard to prove {\em directly} from the definition of the 
edge dependent random graph model.

\end{document}